\documentclass[11pt,twoside]{article}
\usepackage{asp2006}
\usepackage{graphicx}
\begin{document}

\title{Differences between the current solar minimum and earlier minima}
\author{Sarbani Basu}
\affil{Astronomy Department, Yale University, New Haven CT, U.S.A.}
\author{Anne-Marie Broomhall, William J. Chaplin, Yvonne Elsworth}
\affil{School of Physics and Astronomy, University of Birmingham, U.K.}
\author{Stephen Fletcher, Roger New}
\affil{Faculty of Arts, Computing, Engineering and Sciences, Sheffield Hallam University, U.K.}

\begin{abstract}

The Birmingham Solar-Oscillations Network (BiSON) has collected
helioseismic data over three solar cycles. We use these data to
determine how the internal properties of the Sun during this minimum
differ from the previous two minima.  The cycle 24 data show
oscillatory differences with respect to the other two sets, indicating
relatively localized changes in the solar interior.  Analysis of MDI
data from Cycle 23 and Cycle 24 also show significant signs of
differences.

\end{abstract}

\vspace{-0.5cm}
\section{Introduction}
\label{sec:intro}

The minimum before Solar Cycle 24 has been much quieter than many
before, with the lowest sustained 10.7cm radio flux since observation
of this proxy began in 1947.  Other differences have been observed
too, as has been discussed in detail in other articles of this
volume. Most of the discussions concern observations at or above the
solar surface. In this paper we try to determine whether or not there
were changes in the solar interior between this minimum and others
before it using helioseismic data.

The Birmingham Solar-Oscillations Network (BiSON) has been collecting
helioseismic data for over three solar cycles (Broomhall et
al. 2009). No other helioseismic data set spans such a long time
interval.  We analyze data for one-year periods starting 1986/04/01,
1995/11/01 and 2008/05/01 to determine whether there are detectable
differences between the thermal structure of the Sun during these
epochs. These epochs corresponds to the minima of cycles 22, 23 and 24
respectively.

\section{Analysis}
\label{sec:anal}

\begin{figure}[!ht]
\centerline{\hbox{\includegraphics[width = 340pt]{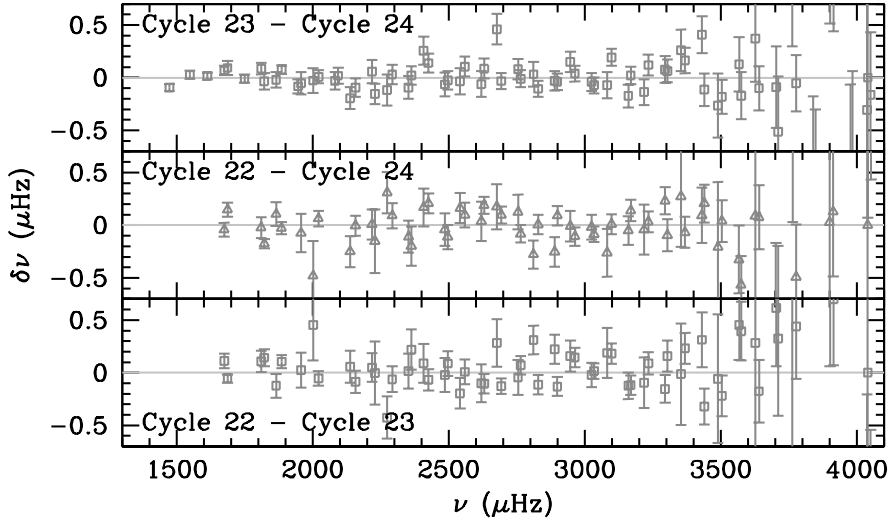}}}
\caption{Differences in low-degree mode frequencies observed for three solar
minima by the BiSON project.
}\label{fig:bdiff}
\end{figure}

The BiSON project makes Sun-as-a-star (i.e., unresolved) observations,
and hence can determine frequencies of only low-degree modes,
principally modes with degree $\ell=0$, 1, 2 and 3. Thus we cannot
invert the frequency differences to determine differences in solar
structure between the three epochs. We therefore, have to use indirect
methods.  We begin by calculating the differences in solar oscillation
frequencies for the three minima. The pair-wise differences are shown
on Fig.~\ref{fig:bdiff}.  As is clear from Fig.~\ref{fig:bdiff}, the
differences are small. However, the differences do not look completely
random --- they show a small frequency dependence. The differences
appear to show a small periodicity as a function of frequency. This
leads us to believe that the frequency differences arise from
differences in deeper layers of the Sun.

\begin{figure}[!ht]
\centerline{\hbox{\includegraphics[width = 340pt]{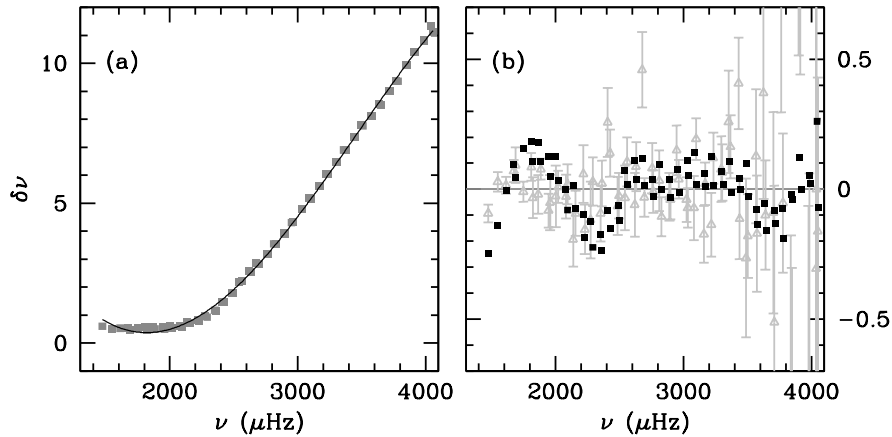}}}
\caption{Panel (a): The difference between the $\ell=0,$ 1, 2, and 3 modes of
solar models BP04 and BSB(GS98) is shown by the grey points. The line in black is a
4th-order polynomial fit to the differences. Panel (b): The black points are the
residuals of removing the 4th-order polynomial from the points in panel (a). The grey points 
with errors bars are the points in the topmost panel of Fig.~\ref{fig:bdiff}.
}\label{fig:mdiff}
\end{figure}

To try and understand the frequency differences we look at the
frequency differences between two solar models. We use model BP04 of
Bahcall et al. (2005) and BSB(GS98) of Bahcall et al.~(2006). These
are two up-to-date solar models. However, they differ in their input
opacities and in the $^{14}N(p,\gamma)^{15}O$ reaction rate. These
give rise to small differences between the models both in the outer
layers and in the core (see Basu et al.~2009). The frequency
differences between the low-degree modes of these two models are shown
in Fig.~\ref{fig:mdiff}(a). As can be seen, the predominant frequency
difference is a smooth function of frequency and is much larger than
the difference in Fig.~\ref{fig:bdiff}. However, the differences in
Fig.~\ref{fig:mdiff}(a) are a hallmark of differences in the very
near-surface layers (see Christensen-Dalsgaard \& Berthomieu 1991).
Frequency differences that arise from deeper differences in structure
can be revealed after subtracting a low-order polynomial in frequency
from the frequency-differences. The residuals following such a removal
are shown in Fig.~\ref{fig:mdiff}(b). As can be seen, the differences
are of the same order as those between the different BiSON sets.
Furthermore the residual differences look very similar to the BiSON
differences. This strengthens our initial conclusion about the
differences in solar structure between the three minima.

Basu et al. (1996) had shown that removing a function of frequency
from the frequency differences is equivalent to filtering out
near-surface information from the set. What remains is information
from the deeper layers. We use that analysis and determine that the
residuals are a result of removing information from layers above
$r\simeq 0.98R_\odot$, i.e., from layers above the second helium
ionization zone.

Basu \& Mandel~(2004) had shown that there are solar-cycle related
changes in the region of the HeII ionization zone. This was confirmed
using BiSON low-degree modes by Verner et al.~(2006). Since the
initial analysis showed differences in the BiSON frequencies arise for
layers at or below the ionization zone, we examine whether we can
detect any such differences between the three solar minima.  Any
spherically symmetric, localized sharp feature or discontinuity in the
Sun's internal structure leaves a definite signature on the solar
p-mode frequencies.  Gough (1990) showed that abrupt changes of this
type contribute a characteristic oscillatory component to the
frequencies $\nu_{n, \ell}$ of those modes that penetrate below the
localized perturbation. The amplitude of the oscillations increases
with increasing ``severity" of the discontinuity, and the wavelength
of the oscillation is essentially the acoustic depth of the sharp
feature. The HeII ionization zone is one such feature, the other being
the base of the convection zone (CZ). The oscillatory signature can be
amplified by taking the second differences of the frequencies, i.e.,
$\delta^2\nu_{n,\ell}=\nu_{n+1,\ell}-2\nu_{n,\ell}+\nu_{n-1,\ell}$.

\begin{figure}[!ht]
\centerline{\hbox{\includegraphics[width = 340pt]{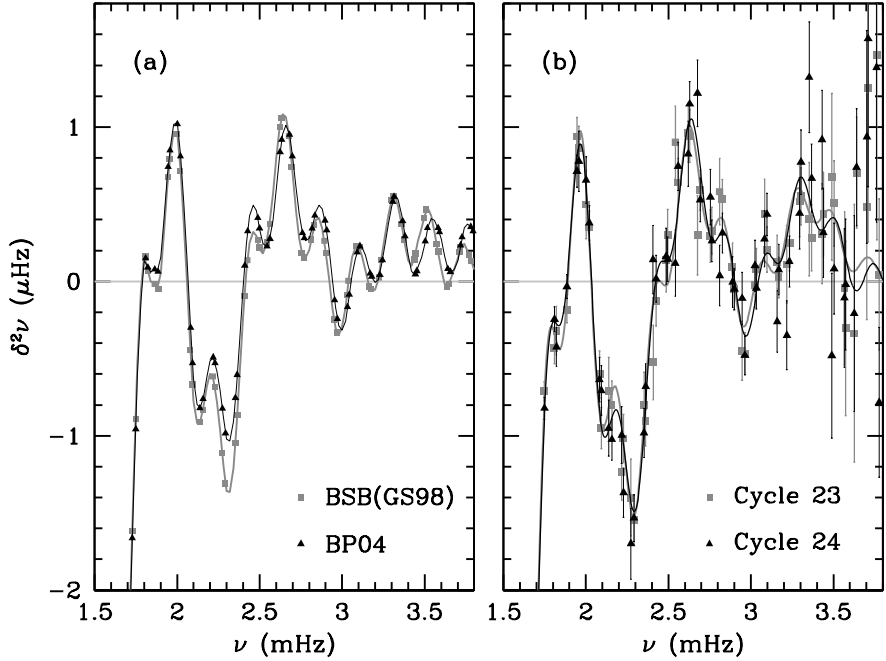}}}
\caption{Panel (a): The second differences of the frequencies of models BP04 and BSB(GS98).
The lines are not fits to the data but have been drawn merely to guide the eye.
Panel (b): The second differences of BiSON sets from the minimum of cycles 23 and 24. Cycle
24 has been omitted for the sake of clarity and also because of the larger errors in that
set. The lines are a fit to the data using the model of Basu et al.~(2004).
}\label{fig:2nd}
\vspace{-0.25 true cm}
\end{figure}

The second differences of the frequencies of the two solar models are
shown in Fig.~\ref{fig:2nd}(a).  Two oscillatory signals are obvious, a
high frequency one that arises from the CZ base, and the low frequency
one from the HeII ionization zone. The signals from the two models are
different, though the differences are small. We, therefore, expect very
small differences between the BiSON data sets.

While the second differences amplify the oscillatory signal, they also
amplify errors. As a result we found that we cannot, at this time, get
reliable results from the frequencies of the minimum of Cycle 22. The
results from Cycles 23 and 24 are shown in Fig.~\ref{fig:2nd}(b).
Also shown are fits to the data using the model of Basu et
al. (2004). The model has three components, an oscillatory part for
the HeII ionization zone, one for the CZ base and a smoother term to
account for near-surface signals.

The amplitudes of the oscillatory terms are frequency dependent and
hence we deal with the average amplitude over the fitting interval of
1.7 mHz to 3.8 mHz. For Cycle 23, we find that the amplitude of the
Helium term is $0.699\pm 0.020\;\mu$Hz, while that for Cycle 24 it is
$0.743\pm 0.020\;\mu$Hz. The differences are not large, but formally
significant at the $2\sigma$ level. If these changes are taken at face
value, it would imply that Cycle 24 has a higher amplitude. This is
consistent with results of Basu \& Mandel~(2004) and Verner et
al.~(2006) who found that the amplitude increases as the level of
activity decreases. This test might therefore be sensitive enough to
distinguish between the small difference of activity between the two
sets. The errors for Cycle 22 are large (mostly because only 3 of the
6 stations in the BiSON network were operating during that period and we
have found several systematic errors that need to be removed from the
data), and we find an amplitude of $0.723 \pm 0.04\; \mu$Hz.  The
wavelength of the HeII signal is a measure of its acoustic depth and
we find a value of $\tau=690.2\pm5.6$s for Cycle 23 and $693\pm 5.5$s
for Cycle 24. The differences are not statistically significant,
however, if we just considered the central vale, the results would
indicate lower sound-speed between the surface and the HeII ionization
zone during the minimum of Cycle 24 compared with Cycle 23.  Cycle 22
has $\tau=690.5\pm 26$s, and the uncertainties are too large to draw
any conclusions.  The differences in the amplitudes and acoustic
depths of the CZ signature is not statistically significant.  Cycles
23 and 24 have amplitudes of $0.283\pm0.042\;\mu$Hz and $0.287\pm
0.042\;\mu$Hz respectively . The acoustic depths are respectively
$\tau=2301\pm15$s and $2317\pm15$s.

As mentioned earlier, the oscillation power spectrum for the minimum
of Cycle 22 shows evidence of some systematic effects, and hence, we
are in the process of re-analyzing the data.

\begin{figure}[!ht]
\centerline{\hbox{\includegraphics[width = 370pt]{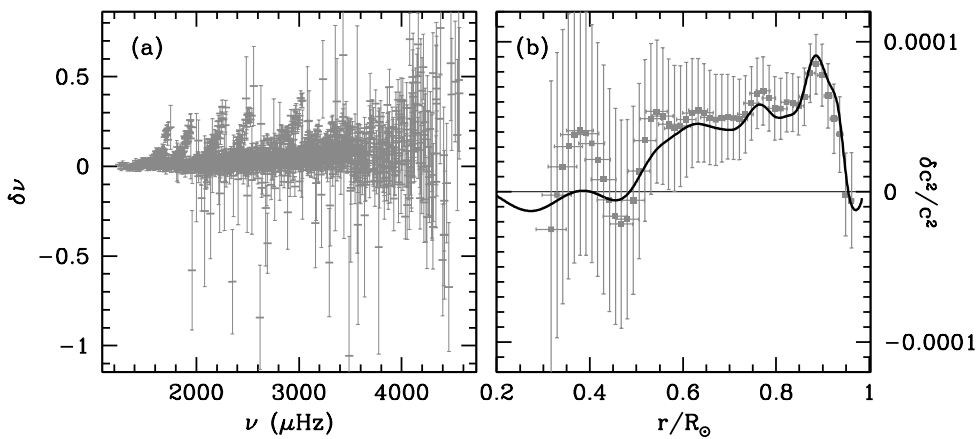}}}
\caption{Panel (a): Frequency differences between MDI 72-day data sets \#1288 (May 1996) and
\#5752 (October 2008). The differences are in the sense (1288$-$5752).
Panel(b): Sound-speed inversion results of the differences in Panel (b). The line
shows the results on an RLS inversion, the points with error bars those of
a SOLA inversion. The results are reliable only where the RLS and SOLA inversions
agree. RLS errors are of the same order as SOLA errors but have not been plotted for the
sake of clarity.
}\label{fig:mdi}
\end{figure}

The Michelson Doppler Imager (MDI) on board {\it SOHO} has observed
helioseismic data for Cycle 23, and hence we also use those data to
determine the differences in the Sun between the minimum of Cycle 23
and the current minimum. These data have p-modes up to $\ell\simeq
200$ and hence we can invert the differences. The frequency
differences between Cycle 23 and 24 are shown in Fig.~\ref{fig:mdi}(a)
and the sound-speed differences obtained by inverting the frequency
differences are shown in Fig.~\ref{fig:mdi}(b).

From Fig.~\ref{fig:mdi}(a) we see that the frequencies for the minimum
of Cycle 23 were higher than those of Cycle 24. This is consistent
with the fact that the minimum of Cycle 23 was more active than that
of Cycle 24.  The frequency differences in Fig.~\ref{fig:mdi}(a) were
inverted using two methods, the Regularized Least Squares (RLS) method
(see Basu \& Thompson 1996) and the Subtractive Locally Optimized
Averages (SOLA) method (Pijpers \& Thompson 1994; Rabello-Soares et
al.~1999). The results are more significant when the two methods give
the same result.  The inversion results show that for the region we
can resolve, the Sun had a somewhat higher sound-speed during the
minimum of Cycle 23 than during the minimum of Cycle 24.

While the differences are small, they do appear to be statistically
significant, at least over a part of the solar interior.  We are
unable to resolve the structure of the outer 10\% due to the lack of
higher-degree modes. The limited number of low-degree modes in the
data sets limits our inversion results in the deeper layers.

\section{Conclusions}

Using data from the Birmingham Solar Oscillation Network and the
Michelson Doppler Imager on board {\it SoHO}, we find evidence that
suggests the structure of the Sun may have been different during the
the minimum of Cycle 24 compared with the minimum of Cycle 23. There
is some evidence that it was different compared to the minimum of
Cycle 22 too, however, Cycle 22 data need to be re-analyzed before we
can make firm conclusions. The evidence points to a lowered
sound-speed during Cycle 24 in layers deeper than 0.98R$_\odot$.

\acknowledgements
This paper utilizes data collected by the Birmingham
Solar-Oscillations Network (BiSON), which is funded by the UK Science
Technology and Facilities Council (STFC). We thank the members of the
BiSON team, colleagues at our host institutes, and all others, past
and present, who have been associated with BiSON.  This paper also
utilizes data from the Solar Michelson Doppler
Imager (MDI) on {\it SoHO}.
SB acknowledges partial support from  NSF grants ATM 0348837 and ATM 073770.


\begin{thebibliography}{}

\bibitem[Basu \& Mandel(2004)]{ba04}
Basu, S., \& Mandel, A. 2004, ApJL, 617, 155

\bibitem[Basu \& Thompson(1996)]{ba96}
Basu, S., \& Thompson, M.J. 1996, A\&A, 305, 631

\bibitem[Basu et al.(2009)]{baetal09}
Basu, S., Chaplin, W.J., Elsworth, Y., New Roger, Serenelli, A.M. 2009, ApJ, 699, 1403

\bibitem[Basu et al.(1996)]{baetal96}
Basu, S., Christensen-Dalsgaard, J., Perez Hernandez, F., \& Thompson, M.J. 1996, MNRAS, 280, 651

\bibitem[Basu et al.(2004)]{baetal04}
Basu, S., Mazumdar, A., Antia, H.M. \& Demarque, P. 2004, MNRAS, 350, 277


\bibitem[Broomhall et al.(2009)]{br09}
Broomhall, A.-M., Chaplin, W.J., Elsworth, Y., Fletcher, S.T., and New, R. 2009, ApJL, 700, 162

\bibitem[Christensen-Dalsgaard \& Berthomieu(1991)]{jcd91}
Christensen-Dalsgaard, J/, \& Berthomieu, G. 1991 in Solar interior and atmosphere, ed. A.N. Cox, W.C. Livingston, M. Matthews
(Tuscon: Univ. of Arizona Press), 401

\bibitem[Gough(1990)]{go90}
Gough, D.O., 1990, in Progress of Seismology of the Sun and Stars, ed. Y. Osaki \& H. Shibahashi (Berlin: Springer), 283 

\bibitem[Pijpers \& Thompson(19994)]{pij94}
Pijpers, F.P., \& Thompson, M.J. 1994, A\&A, 281, 231

\bibitem[Rabello-Soares et al.(1999)]{ra99}
Rabello-Soares, M.C., Basu,. S., \& Christensen-Dalsgaard, J. 1999, 309, 35

\bibitem[Verner et al.(2006)]{ver06}
Verner, G.A., Chaplin, W.J., \& Elsworth, Y. 2006, ApJL, 640, 95

\end{thebibliography}
\end{document}